\documentclass[aps,prb,floatfix,twocolumn,superscriptaddress]{revtex4-1}
\usepackage{graphicx,amsmath,amssymb,subfigure,epsfig,psfrag,color,verbatim,array}

\newcommand{\myhrulefill}{}

\newcommand{\methoda}{\begin{equation} \tau = \frac{2-\alpha}{\beta + \gamma}+1 \end{equation}}
\newcommand{\methodaa}{\begin{equation} \tau = \frac{2-\alpha}{2-\beta-\alpha}+1\end{equation} because $\beta$ is so small this equation is relatively insensitive to $\alpha$}

\begin{document}
\title{Spatial Complexity Due to Incipient Electronic Nematicity in Cuprates}
\author{B.~Phillabaum}
\affiliation{Department of Physics, Purdue University, West Lafayette, IN  47907}
\author{E.~W.~Carlson}
\affiliation{Department of Physics, Purdue University, West Lafayette, IN  47907}
\author{K.~A.~Dahmen}
\affiliation{Department of Physics, University of Illinois, Urbana-Champaign, IL 61801}
\date{\today}

\begin{abstract}
Surface probes such as scanning tunneling microscopy (STM)
have detected complex patterns at the nanoscale,
indicative of electronic inhomogeneity, in a variety
of high temperature superconductors.  
In cuprates, the pattern formation is associated with
the pseudogap phase, a precursor to the high temperature
superconducting state.  
Symmetry breaking (i.e. from C4 to C2) in the form of
electronic nematicity has recently been
implicated as a unifying theme of the pseudogap phase,\cite{Lawler:2010p639}
however the fundamental physics governing the nanoscale
pattern formation has not yet been identified.  
Here we use universal cluster properties extracted from STM studies of
cuprate superconductors
in order to identify the fundamental physics controlling the complex pattern formation.
We find that the pattern formation is set
by a delicate balance between disorder and interactions,
leading to a fractal nature of the cluster pattern.
The method introduced here may be extended to a variety of surface probes,
enabling the direct measurement of the {\em dimension} of the phenomenon
being studied, in order to determine whether the phenomenon arises from the
bulk of the material, or whether it is confined to the surface.
\end{abstract}

\maketitle

While lanthanum-based cuprate superconductors display striking evidence
of rotational symmetry breaking in the pseudogap
regime, sometimes even leading to translational symmetry
breaking\cite{Tranquada:2004p132},
such issues have been harder to settle in the
higher transition temperature compounds such as YBCO and BSCCO.
Recent experimental progress on YBCO has shed new light,
including nematic behavior in the pseudogap regime
detected via Nernst effect\cite{Daou:2010p784},
transport\cite{Ando:2002p780}, and neutron scattering\cite{Hinkov:2008p878},
as well as evidence of time-reversal symmetry breaking
detected via neutron scattering\cite{Fauque:2006p423} and the
Kerr effect\cite{Xia:2008p306}.
The detection of a glass of unidirectional domains via STM in 
NCCOC and Dy-Bi2212 has now been followed
by the dramatic demonstration of rather large electron nematic
domains in BSCCO\cite{Lawler:2010p639,Kivelson:1998p645}.

Because of the fragile nature of rotational symmetry breaking,
certain classes of disorder forbid long-range orientational order in 2D systems,
and significantly suppress the order in strongly layered 
systems.\cite{Carlson:2006p651,Zachar:2003p207}  
It is therefore critical to understand the effects 
of both quenched disorder and interactions between
nematic domains in these systems.  
As shown in Ref.~\onlinecite{Carlson:2006p651},
the orientational degree of
freedom of the  electron nematic in the cuprates 
in the presence of quenched disorder maps
to a disordered Ising model in the following way.
When an electron nematic forms, there is a preferred
orientation to the electronic degrees of freedom.
We consider Cu-O planes which are locally $C_4$
symmetric, with an incipient elelctron nematic
which breaks the local rotational symmetry 
of the host crystal from $C_4$ to $C_2$,
leading to two possible nematic orientations.
We coarse grain the system,
and define a local nematic order parameter by 
$\sigma = \pm 1$, corresponding to the two allowed
orientations.    
The tendency for neighboring nematic regions to
align is modeled as a ferromagnetic nearest-neighbor
interaction.

Material disorder in the form of, {\em e.g.}, dopant atoms
competes with the ferromagnetic coupling between 
local nematic directors.  
Upon coarse graining, there
are two broad classes of disorder which present themselves
at the order parameter level:  local energy density disorder
(which includes random $T_c$ disorder), and random field disorder.
Local density disorder may arise in the form of, {\em e.g.},
random bond disorder, in which the strength of the ferromagnetic
coupling varies from coarse-grained site to coarse-grained site in the system.
In addition, the local amplitude of the nematic order parameter
can vary spatially\cite{Lawler:2010p639}.  In an order parameter description, 
this type of disorder may be subsumed into randomness in the bond strengths.  
The other class of disorder, random field disorder, 
arises when the pattern of doping atoms breaks rotational symmetry,
thus favoring one or the other orientation of the nematic director in that region.
This type of disorder couples linearly to the nematic director.  
The purpose of this paper
is to gain an understanding, at the order parameter level,
of both the relative importance of interactions and disorder,
and also to discern the type of disorder, thus arriving at
an order parameter level description of the fundamental
physics controlling the inhomogeneous pattern formation
observed at the nanoscale in these materials.  
We consider a general model encompassing both classes of disorder:
\begin{eqnarray}
H = &-& \sum_{\left< i j \right>_{||}} J^{||}(1+\delta J^{||}_{ij})\sigma_i \sigma_j  \\ \nonumber
&-& \sum_{\left< i j \right>_{\perp}} J^{\perp}(1+\delta J^{\perp}_{ij})\sigma_i \sigma_j
- \sum_i(h+ h_i) \sigma_i~.
\label{eqn:rfim}
\end{eqnarray}
Here, $J^{||}$ sets the overall strength of the in-plane ferromagnetic
coupling between nearest neighbor Ising nematic variables,
is an in-plane ferromagnetic coupling between
nearest neighbor Ising nematic variables, and $\delta J^{||}_{ij}$
represents bond disorder in the coupling strength.
$J^{\perp}$ represents the overall coupling strength between
Ising variables in neighboring planes, and 
$\delta J^{\perp}_{ij}$ is bond disorder in the interplane coupling strength.
The random field $h_i$ is chosen from a gaussian probability
distribution centered about zero,
with width $\Delta$, which we call the ``random field strength''.
We have chosen to discretize the system on a cubic lattice
for simplicity, a choice  which does not affect the 
universal properties of the model 
as long as the mean value of the random bond couplings remains positive.

The field $h = h_{\rm int} + h_{\rm ext}$
represents an orienting field which breaks rotational symmetry.
The external contribution $h_{\rm ext}$
may be achieved by the application of, {\em e.g.},
magnetic fields, uniaxial pressure, or high currents among others.
In the systems we have analyzed, data was taken in the
absence of applied fields.    Another source of finite $h$ may be
internal crystal effects $h_{\rm int}$, such as the chains in YBCO may present.
Such issues do not arise in NCCOC or BSCCO,
and it is appropriate in both NCCOC and BSCCO to 
set the internal field $h_{\rm int}=0$ in our model.

The order parameter
$m = {1/N} \sum_i \sigma_i$ describes the degree of
orientational order in the system. 
In the limit of zero random field strength and no bond disorder,
the model has a finite-temperature continuous phase transition for
dimension $d \ge 2$, from a disordered phase 
to an ordered electron nematic, {\em i.e.} with long range orientational order $<m> \ne 0$.
However, 
for any finite random field strength, long range orientational
order is {\em forbidden} in two dimensions.
In three dimensions, there is a finite critical disorder strength of the random field,
$\Delta_c^{\rm 3D} = 2.27 J_{\parallel}$.\cite{Middleton:2002p600}
In strongly layered systems, the critical disorder strength is finite, but significantly reduced
from the 3D value.\cite{Zachar:2003p207}   
In contrast with random field models, 
weak random bond disorder does not forbid nematic order in two dimensions;
rather, the phase transition is governed by clean Ising model exponents.
In layered and  three dimensional systems, the phase transition 
of the random bond model is controlled by the disordered fixed point ``R''.    
In the presence of both random bond and random field disorder,
the universality class is that of the random field model.

\begin{figure}[thb]
\centering
	\includegraphics[width=0.75\columnwidth]{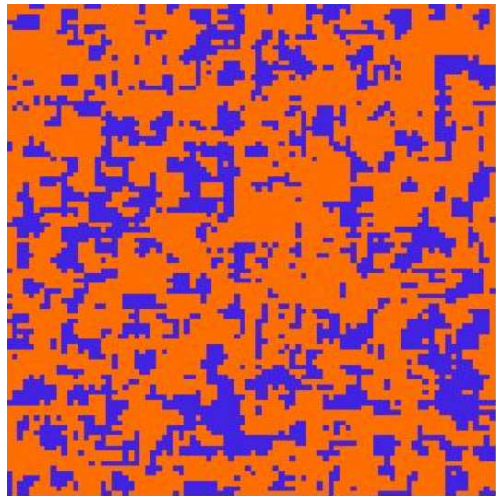}
\caption{
Extraction of Ising domains from tunneling asymmetry maps in 
Dy-Bi2212, with $T_c \approx 45$K, taken at $T = 4.2K$.
Notice the large, percolating cluster (orange),
with isolated flipped domains inside.  
\label{fig:DBSCCO45}
}
\end{figure}

\begin{table*}
\begin{centering}
\vspace{.1in}
\begin{tabular}{|c|c|c|c|}
\hline
{}&{Clean Ising Model}&{Random Bond Disorder}&{Random Field Disorder}\\
\hline
\hline
{$\tau$}&{2.067 \cite{cardy-book,loh-database}}&{2.067 \cite{cardy-book,loh-database}}&{1.6\cite{Alava:1998p822}}
\\ \hline
{$d_c$}&{1.125\cite{Middleton:2002p600,chaikin-lubensky}}&{1.125\cite{Middleton:2002p600,chaikin-lubensky}}&{1.9\cite{Seppala:1998p823}}
\\ \hline
{$d-2+\eta$}&{0.25\cite{chaikin-lubensky}}&{0.25\cite{chaikin-lubensky}}&{1\cite{Bray:1985p580}}
\\ \hline
\end{tabular}	
\caption{{\bf Theoretical Critical Cluster Exponents of Two Dimensional Ising Models.}
In two dimensions, random bond disorder is irrelevant, and the order to disorder
phase transition is controlled by the clean Ising model critical point.
For the random field Ising model in two dimensions, long range order is forbidden at finite disorder strength.
However, there is an unstable fixed point at T=0 and zero disorder.  The table reports  values from the literature for this point, since it may affect the scaling in some regimes.
}
\label{tab:2D-Ising}
\end{centering}
\vspace{-10pt}
\end{table*}
\begin{table*}
\begin{centering}
\vspace{.2in}
\vspace{.1in}
\begin{tabular}{|c|c|c|c|}
\hline
{}&{Clean Ising Model}&{Random Bond Disorder}&{Random Field Disorder}
\\ \hline \hline
{$\tau$}&{2.21 \cite{cardy-book,loh-database}}&{2.221 \cite{cardy-book,Xiong:2010p848}}&{2.02$\pm$0.03\cite{liu-private,Liu:2009p873}}
\\ \hline
{$d_c$}&{2.11\cite{Middleton:2002p600,chaikin-lubensky}}&{1.98\cite{Middleton:2002p600,Xiong:2010p848}}&{2.78 $\pm$ 0.05\cite{Liu:2009p838,Liu:2009p873,PerezReche:2003p846}}
\\ \hline
 {$d-2+\eta$}&{1.04\cite{chaikin-lubensky}}&{1.06\cite{Xiong:2010p848}}&{1.5$\pm$0.05\cite{Rieger:1995p850}}
 \\ \hline
\end{tabular}
\caption{{\bf  Theoretical Critical Cluster Exponents of Three Dimensional Ising Models.}
}
\label{tab:3D-Ising}
\end{centering}
\vspace{-10pt}
\end{table*}

\begin{table}
\begin{centering}
\begin{tabular}{|c|c|}
\hline
{}&{Dy-Bi2212 at T=4.2K}\\
\hline
\hline
{$\tau$}&{1.71 $\pm$ 0.07}
\\ \hline
{$d_c$}&{1.56 $\pm$ 0.02}
\\ \hline
{$d-2+\eta$}&{0.26 $\pm$ 0.03}
\\ \hline
\end{tabular}
\caption{{\bf Critical Cluster Exponents from STM on 
the cuprate superconductor Dy-Bi2212.}}
\label{tab:extracted-exponents}
\end{centering}
\end{table}

\section{Universal Spatial Properties of Clusters}
Fig.~\ref{fig:DBSCCO45}
shows our extraction of locally oriented (nematic) domains from scanning tunneling microscopy data on 
Bi$_2$Sr$_2$Dy$_{0.2}$Ca$_{0.8}$Cu$_2$O$_{8+\delta}$ (Dy-Bi2212)
from Fig.~S3 of Ref.~\cite{Kohsaka:2007p713}, reported as an ``R-map,'' where at each position $\vec{r}$,
$R(\vec{r},V) = I(\vec{r},+V)/I(\vec{r},-V)$ is the ratio of the tunneling current at positive voltage to 
that at negative voltage.\cite{Kohsaka:2007p713,Anderson:2006p853,Randeria:2005p854}  
Based on the ``R''-map, the presence of local, unidirectional domains of 
width $4a_o$ were noted\cite{Kohsaka:2007p713}, corresponding to the
distance between ``legs'' of the $4 a_o$-wide ladders, where
$a_o$ is the Cu-Cu distance within the Cu-O planes.
There is also
a coexisting local density wave at $a_o$, corresponding to the
distance between ``rungs'' on the ladders.  
To effect the mapping to Ising nematic domains,
we take a local spatial Fourier transform (FT) of the R-map in any given region, and then 
focus on the $a_o$ local density wave.  
There are two ``flavors'' to this density wave, either oriented along the $a$ axis,
or along the $b$ axis, depending on the local orientation of the ladders.  The relative weight of the two density waves can be discerned
from comparing the $a_o$ periodicity peaks in the local FT,
comparing the weight of the $a_o$ peak in the $a$ direction to that in the $b$ direction.
We assign a local Ising variable based on this relative weight.
We have checked that our 
results are insensitive to details such as
the size of the FT window and the Ising lattice spacing.  
\footnote{Different techniques have been used in Ref.~\cite{Lawler:2010p639}
to extract a nematic order parameter associated with symmetry breaking {\em within}
the crystal unit cell.  Our method focuses on nematicity associated with the $a_o$
periodicity itself, and has the advantage that it does {\em not} depend sensitively on the phase
of the complex Fourier transform.  For example, it is not necessary with our method to have
detailed information about the exact location of each atom.}
Note that our assignment of the Ising cluster map is independent 
of the local strength of the nematicity,
which can be subsumed into
randomness in the Ising couplings $J_{ij}$, contributing to random bond disorder.

\begin{figure}[thb]
\centering
	\includegraphics[width=\columnwidth]{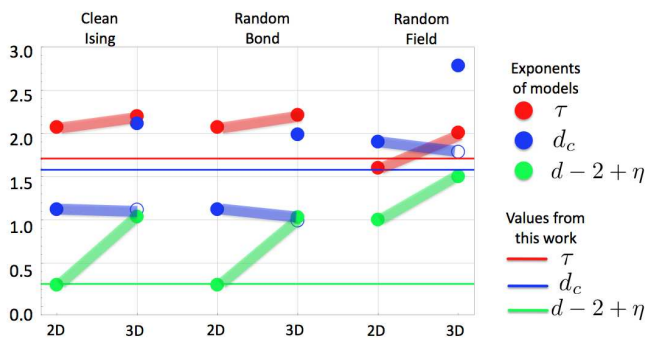}
\caption{
Chart comparison of our results for Dy-Bi2212 compared to the theoretical models. 
Solid circles represent the values reported in the literature, summarized in Tables~\ref{tab:2D-Ising}
and \ref{tab:3D-Ising}.  Red circles represent $\tau$, blue circles represent $d_c$,
and green circles represent $d-2+\eta$.   In comparing our results to the 3D models, we have assumed 
that at the surface, one observes 
a two-dimensional cross section of a 3D cluster, and we subtract 1 from the fractal dimension $d_c$.  These values are represented by the open blue circles.  Thick lines connecting circles
represent putative crossover of exponents from 2D to 3D behavior in a 
layered material.  The  horizontal lines in red, blue, and green are our results for the spatial characteristics of clusters in Dy-Bi2212 as reported in Fig. S3 of Ref.~\onlinecite{seamus-glass}
\label{fig:Dy-Bi2212-chart}
}
\end{figure}

The extracted cluster maps resemble
cluster patterns observed in numerical studies of Eqn.~\ref{eqn:rfim},\cite{Middleton:2002p600}
consistent with the idea of mapping disordered nematics
to a disordered Ising model.\cite{Carlson:2006p651}
We show in Tables~\ref{tab:2D-Ising} and \ref{tab:3D-Ising}
cluster properties  which can be extracted from 
the spatial configurations of clusters, as defined in the following references:
 $\eta$ is the spin-spin correlation function exponent\cite{Bray:1985p580}, 
$\tau$ is the cluster size distribution exponent\cite{Perkovic:1999p3}, and
$d_c$ is the fractal dimension of the spanning cluster\cite{Middleton:2002p600}.
The theoretical values available from  the literature are shown for clean Ising models,
random bond Ising  models, and random field Ising models in both
two and three dimensions,\cite{Alava:1998p822,Seppala:1998p823,Picco:2006p869}
or derived from a combination of theoretical exponents available in the
literature and scaling relations.  
(See the Appendix.)
Notice how {\em distinct} the numbers are from one universality class to the next.
This implies that different universality classes are sufficiently different
that one can in principle use these methods to identify the universality class
controlling the inhomogeneous pattern formation.
In particular, because critical exponents are so sensitive to dimension, 
this method can in principle be used to allow any surface
probe to {\em directly measure the dimension} of the phenomenon 
studied, to determine whether the phenomenon is happening only on the 
surface of the material ($d=2$), or whether it arises from the bulk of the material
($d=3$).

We  apply quantitative cluster analysis methods\cite{Middleton:2002p600,Perkovic:1999p3,Liu:2009p838,Liu:2009p873,PerezReche:2003p846}
from the statistical mechanics of disordered systems to identify the fundamental physics controlling the inhomogeneous pattern formation.
Our results using published
STM data\cite{Kohsaka:2007p713} on Dy-Bi2212, 
a material which displays evidence of a disordered electron nematic at the surface,\cite{Kohsaka:2007p713,Lawler:2010p639}
are shown in Table~\ref{tab:extracted-exponents}.
Fig.~\ref{fig:Dy-Bi2212-chart} charts a direct comparison between
our values extracted from data, and the theoretical fixed point exponents of disordered Ising models.  
Because of the strongly layered nature of Dy-Bi2212, we expect there to be
a regime of scaling which follows the 2D universality class,
which then crosses over to the 3D universality class at longer length scales.\cite{Zachar:2003p207}

\subsection{Cluster Size Distribution}
\begin{figure}
\includegraphics[width=\columnwidth]{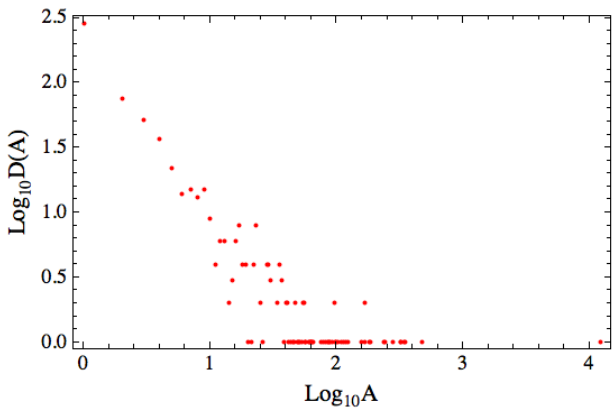}
\caption{Raw cluster size distribution.}
\label{fig:rawtau}
\end{figure}

\begin{figure}
\includegraphics[width=\columnwidth]{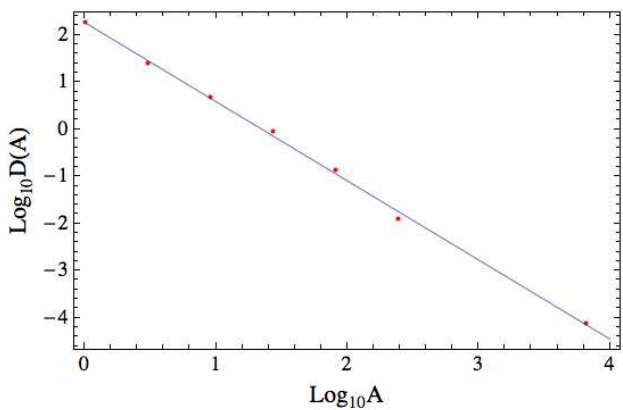}
\caption{Cluster Size Distribution after logarithmic binning,
used to calculate the critical exponent $\tau$.}
\label{fig:binnedtau}
\end{figure}

The exponent $\tau$ characterizes the distribution of cluster sizes.
Fig.~\ref{fig:DBSCCO45} shows the clusters identified from Dy-Bi2212.
Here, a  single cluster is a nearest-neighbor set of like-oriented Ising variables
on the square lattice we have chosen to use. 
Fig.~\ref{fig:rawtau} shows the cluster size distribution,
a histogram of cluster sizes.    The results are noisier at 
large cluster size, due to the finite field of view (FOV).  
Fig.\ref{fig:binnedtau} shows the same distribution
with logarithmic binning.    
Near a critical point, the cluster size distribution is of the form
$D(A) \propto A^{-\tau}$.  Using a straightforward fit to this form,
we find $\tau = 1.71 \pm 0.07$.  
There is one large spanning cluster, represented by the 
last point in Fig.~\ref{fig:binnedtau}.  
Near criticality, scaling can be observed over a finite range.
Note that in this case, the spanning cluster is also in the regime of scaling,
and almost 4 decades of scaling are present in this cluster property,
and this extraction of the exponent $\tau$ can be considered
quite reliable.

\subsection{Fractal Dimension of Clusters}

\begin{figure}
\includegraphics[width=\columnwidth]{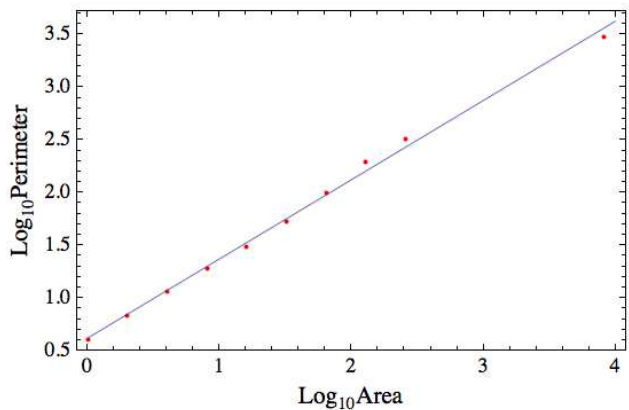}
\caption{The fractal dimension of clusters, $d_c$, 
relates the perimeter of clusters to their area.  (See text for details.)
Logarithmic binning has been used.}
\label{fig:dsc}
\end{figure}

The fractal dimension of clusters relates the 
surface area of each cluster to its volume.  
Because STM is a surface probe, the available information
is two-dimensional, thus the fractal dimension of clusters
relates their perimeter $p$ to the area $a$ of each cluster:
$p \propto a^{d_c/d}$.\cite{Middleton:2002p600}  
Using $d=2$ because only a two dimensional cross section
of the cluster properties is available, 
a straightforward fit  gives $d_c = 1.56 \pm 0.02$.
It is evident from Fig.~\ref{fig:dsc} that 
the spanning cluster is also in the scaling regime for this measure,
leading to almost 4 decades of scaling for the fractal dimension,
and as with $\tau$, this extraction of the exponent $d_c$ can also be
considered quite reliable.

\subsection{Spatial Correlations}
\begin{figure}
\includegraphics[width=\columnwidth]{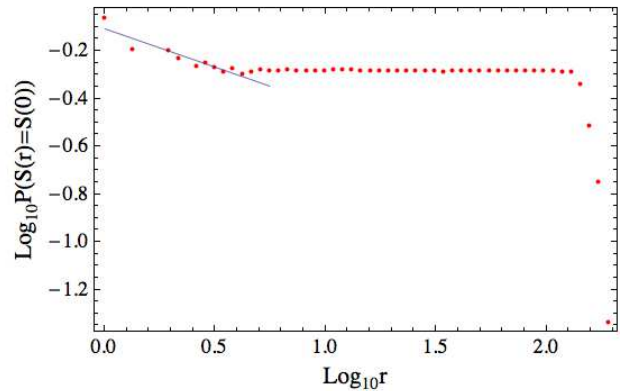}
\caption{Logarithmically binned data for calculating the critical exponent $\eta$.}
\label{fig:eta}
\end{figure}

Within the Ising description, the spin-spin correlation function
of the Ising pseudospin variables is described by
$G(r) \propto |r|^{-(d-2+\eta)}$.  
Within the field of view available, we have averaged over all sites
to obtain the spin-spin correlation function plotted in 
Fig.~\ref{fig:eta}.  
At short distances, a weak region of scaling is evident,
which extends roughly half of a decade.  
With only a half decade of scaling, this measure should 
be considered {\em unreliable}, especially since the presence
of an as-yet undetermined scaling function may skew results
at short distances.  Nevertheless, for completeness,
we note that the value of $d-2+\eta$ obtained from this simple fitting procedure is 
$d-2+\eta = 0.26\pm 0.03$.

\section{Discussion}
The value of the cluster size distribution exponent $\tau$ 
displayed by the clusters in Dy-Bi2212 lies between the 2D and 3D values of the RFIM,
indicating that the cluster size distribution is consistent with the scaling of a 
{\em layered} RFIM.  By contrast, the value of $\tau$ is $17\%$ away from the
clean and random bond values denoted by thick red lines in Fig.~\ref{fig:Dy-Bi2212-chart}.  
Furthermore, the data show almost 4 decades of scaling for $\tau$
indicating the value of $\tau$
extracted is highly reliable.  
In comparing the fractal dimension of clusters $d_c$ to 3D models, we make the ansatz
that we have analyzed a 2D cross section of a 3D cluster,
and subtract $1$ from the expected value of $d_c$.
We see then that the value of $d_c$ extracted from experiment
is close to the theoretical RFIM values (within $12\%$),
whereas it is $39\%$ away from the random bond and clean models.
There are almost 4 decades of scaling for the fractal dimension $d_c$,
making this a  reliable extraction of this universal
cluster property.  
The value of $d-2+\eta$ extracted from a simple fit to the spin-spin correlation
function is inconsistent with the random field universality classes,
but it is consistent with Ising models, either clean or random bond models close to two dimensions.
However, this is the least reliable exponent extracted, as it displays less than one decade of scaling. 
The most reliable properties we have extracted
are $\tau$ and $d_c$ with almost 4 decades of scaling including the spanning cluster.
(We note that even without the spanning cluster, over 2 decades
of scaling are present in both of these quantities.)
Overall, there is a striking similarity 
between the spatial properties of clusters observed in STM
and universal cluster properties of disordered Ising models in 2D and 3D,
particularly of the random field type.

The presence of power law scaling in these
quantities suggests that the pattern formation in Dy-Bi2212 is near criticality.  
Because the critical region is large in the RFIM
it is possible to observe the scaling in a broad range
of parameters near the critical point.\cite{Perkovic:1995p2}
For example, two decades of scaling appear in the
avalanche size distribution of the nonequilibrium 3D RFIM
even at a disorder strength of $\Delta=4J$, 
which is $85\%$ larger than the nonequilibrium 
critical disorder strength $\Delta_c^{\rm 3D} = 2.16J$.\cite{Perkovic:1995p2}
The presence of power law scaling is also indicative of the fractal
nature of this pattern formation, suggesting a potential connection to recent work
on the fractal nature of doping in some high temperature superconductors.\cite{bianconi-fractal-doping}
 
In a true 2D condensed matter system ({\em e.g.} , with a finite disorder concentration),
random field disorder would forbid long-range order of an electron nematic.
However, in any finite size 2D system, the system can {\em appear} to be ordered whenever
the correlation length $\xi$ is larger than the system size observed. 
Although the material we have considered is strongly layered, the layered RFIM
ultimately  flows to the 3D universality class,\cite{Zachar:2003p207} which has a finite critical disorder strength.
Because of the weak coupling between planes, the critical 
disorder strength is suppressed from the
3D limit, {\em i.e.} $R_c(J_{\perp} < J_{\parallel} < R_c^{3D}$).
Based on this, coupled with the fact that 
long-range-ordered nematicity has not been experimentally
detected in BSCCO via a bulk probe, it is reasonable to expect 
the disorder strength in this material 
to be beyond the critical disorder strength.
However, there is a spanning cluster in Fig.~\ref{fig:DBSCCO45},
so the system cannot be too far above the critical disorder strength,
since deep in the disordered phase, clusters are small. 
A spanning cluster was also observed in Ref.~\onlinecite{Lawler:2010p639}.
We thus conclude that this material is in the {\em intermediate disorder
regime},\cite{Zachar:2003p207} where the disorder strength is small within a plane, but
strong between planes,  $J_{||}  \gg \Delta >> J_{\perp}$.
Although long-range nematic order is not possible in this regime,
clusters within each plane can have a long correlation length.
\vspace{.1in}

\myhrulefill\section{Conclusions}
In conclusion, we have used quantitative cluster methods from the statistical mechanics of
disordered systems to identify the fundamental physics controlling the 
complex pattern formation revealed by STM in Dy-Bi2212.  
We find that the patterns are scale-invariant and fractal, indicating that
Dy-Bi2212 is near a critical point, due to the interplay of
local rotational symmetry breaking (electronic nematicity)
with quenched disorder.
We furthermore identify this critical point as being in the Ising universality class,
with random field disorder the dominant type of disorder.
We find that the disorder strength is in the {\em intermediate} regime,
with disorder being a weak effect within each plane (leading to large
clusters within a plane), while disorder is a strong effect
in the interplane direction (leading to a lack of true long range order),
indicating the pattern formation may best be described by 
a layered random field Ising model.  
Furthermore, this method may be extended to other materials and surface probes.
For spatial patterns that may be tuned through or sufficiently near a critical point,
this allows the determination of the dimension of the phenomenon being studied,
enabling surface probes to distinguish whether the pattern formation is merely
on the surface, or extends throughout the bulk of the material.  

\myhrulefill\section{Acknowledgments}
It is a pleasure to thank B. Brinkman, E. Fradkin and S. Kivelson for conversations.
This work was supported by Research Corporation 
and by NSF Grant Nos. DMR 08-04748, DMR 10-05209, and DMR 03-25939 ITR (MCC).
EWC thanks ESPCI for hospitality.  

\section*{Appendix}
\subsection{Theoretical Values of Critical Exponents}
The exact critical exponents we require in order  to compare to known theoretical results of 
Ising models are not necessarily directly in the literature.  However, scaling relations allow us
to infer values for the cluster size distribution exponent, $\tau$, and the fractal dimension of 
clusters, $d_c$.    
To infer $d_c$, we employ the scaling relation\cite{Middleton:2002p600}:
\begin{equation}
d_c = {\alpha - 1 \over \nu} + d + {\beta \over \nu}
\end{equation}
To infer $\tau$ we used the scaling relation \cite{cardy-book}
\methoda

along with the Rushbrooke identity\cite{cardy-book} to eliminate $\gamma$ in favor of $\alpha$:
\begin{equation}
\alpha+2\beta+\gamma = 2 \rightarrow \gamma = 2-2\beta-\alpha
\end{equation}

to get the final expression:
\methodaa

Some of these exponents were extracted directly from the literature, in order  to avoid confusion an asterisk will be used to denote that a particular value was inferred via relations rather than directly reported in the literature.

\begin{table}[htb]
\begin{centering}
\begin{tabular}{|c|c|c|}
\hline
Exponent & 2D Clean Ising Model & 3D Clean Ising Model \\ \hline
\hline
$d-2+\eta$  &  1/4 \cite{Kardar} &1.04\cite{chaikin-lubensky} \\
$\alpha$& 0 \cite{Kardar}&0.1118(6) \cite{chaikin-lubensky}\\
$\beta$&1/8\cite{Kardar}&0.326(4) \cite{chaikin-lubensky}\\
$\nu$ &1\cite{Kardar}&0.6294(2)\cite{chaikin-lubensky}\\
$\tau$&$31/15$*&2.21*\\
$d_c$&$9/8$*&2.11*\\
\hline
\end{tabular}
\caption{Theoretical values of critical exponents
in 2D and 3D clean Ising models.}
\label{tab:clean-exponents}
\end{centering}
\end{table}
\begin{table}[htb]
\begin{centering}
\vspace{.1in}
\begin{tabular}{|c|c|}
\hline
Exponent & 3D RBIM ``R'' point \\ \hline
$d-2+\eta$  & 1.029\cite{Berche:2004p933} \\
$\alpha$&-0.04\cite{Berche:2004p933}\\
$\beta$&0.35\cite{Berche:2004p933}\\
$\gamma$&1.34\cite{Berche:2004p933}\\
$\nu$&0.68\cite{Berche:2004p933}\\
$\tau$&2.21* \\
$d_c$&1.99*\\
\hline
\end{tabular}\\
\caption{Theoretical values of critical exponents
in 3D Ising models with weak random bond disorder.
Note that transition in the 2D random bond Ising  model is controlled by 
the clean fixed point.}
\end{centering}
\end{table}
\begin{table}[htb]
\begin{centering}
\begin{tabular}{|c|c|c|}
\hline
Exponent & 2D RFIM & 3D RFIM \\ \hline
$d-2+\eta$  &  1 \cite{Bray:1985p580} &$0.5 \pm 1.03$\cite{Hartmann:2001p872} \\
$\alpha$&&0.63 \cite{Hartmann:2001p872}\\
$\beta$&&0.017(5) \cite{Middleton:2002p600}\\
$\nu$&&1.37(9)\cite{Middleton:2002p600}\\
$\tau$&1.6\cite{Alava:1998p822}&2.01*\\
$d_c$&1.9 \cite{Seppala:1998p823}&2.74*\\
\hline
\end{tabular}
\caption{Theoretical values of critical exponents in 2D and 3D random field Ising models.}
\end{centering}
\end{table}

\subsection{Harris criterion considerations}
The Harris criterion states that local energy density disorder (such as random $T_c$ disorder
and random bond disorder)
is irrelevant if $d\nu >2$, where the exponents refer to the clean model.  
In the presence of hyperscaling (obeyed by the clean and random bond
cases), $d \nu = 2-\alpha$ implies that the Harris criterion reduces to $\alpha < 0$.  
In the 3D clean Ising model, $\alpha \approx 0.1$\cite{cardy-book}, and randomness is relevant.
In the 2D clean Ising model, $\alpha = 0$ and such randomness is marginal.  
For the 3D case with weak bond disorder, there is a disordered fixed point with new exponents.
In the 2D case with weak bond disorder, it has been shown that the system flows toward the clean model.\cite{Shalaev:1994p934,Picco:2006p869}.

\bibliographystyle{forprb} 
\bibliography{rfim,ecrfim,detectingnematics,rfim-frompapers} 

\end{document}